\documentclass[%
 reprint,
 amsmath,amssymb,
 aps,
]{revtex4-2}

\usepackage{graphicx}
\usepackage{dcolumn}
\usepackage{bm}

\usepackage{graphicx}
\usepackage{epstopdf, epsfig}
\usepackage{lipsum}
\usepackage{amsmath,amssymb}
\usepackage{amsfonts}
\usepackage{graphicx}
\usepackage{todonotes}
\usepackage{tikz}
\usepackage{color}
\usetikzlibrary{shapes}

\definecolor{MATLABcyan}{RGB}{77, 190, 238}
\definecolor{MATLABred}{RGB}{217, 83, 25}
\definecolor{MATLABgreen}{RGB}{119, 172, 48}
\definecolor{MATLAByellow}{RGB}{237, 177, 32}
\definecolor{MATLABpurple}{RGB}{126, 47, 142}
\definecolor{MATLABblue}{RGB}{0, 114, 189}
\definecolor{MATLABgrey}{RGB}{128, 128, 128}

\begin{document}

\preprint{APS/123-QED}

\title[]{Tunneling in a Lorenz-like model for an active wave-particle entity}

\author{Runze Xu$^{1}$}
\author{Rahil N. Valani$^{1,2}$}\email{rahil.valani@physics.ox.ac.uk}
\affiliation{$^1$School of Computer and Mathematical Sciences, University of Adelaide, South Australia 5005, Australia}
\affiliation{$^2$Rudolf Peierls Centre for Theoretical Physics, Parks Road,
University of Oxford, OX1 3PU, United Kingdom}


\date{\today}

\begin{abstract}

Active wave-particle entities (WPEs) emerge as self-propelled oil droplets on the free surface of a vibrating oil bath. The particle (droplet) periodically imprints decaying waves on the liquid surface which in turn guide the particle motion, resulting in a two-way coupling between the particle and its self-generated waves. Such WPEs have been shown to exhibit hydrodynamic analogs of various quantum features. In this work, we theoretically and numerically explore a dynamical analog of tunneling by considering a simple setup of a one-dimensional WPE incident on an isolated Gaussian potential barrier. Our idealized model takes the form of a perturbed Lorenz system which we use to explore the dynamics and statistics of barrier crossing as a function of initial conditions and system parameters. Our work highlights that velocity fluctuations of the WPE at high memories that are rooted in non-equilibrium features of the Lorenz system, such as spiraling motion towards equilibrium points and transient chaos, give rise to - (i) sensitivity and unpredictability in barrier crossing, (ii) smooth variations in transmission probability as a function of system parameters, and (iii) wave-like features in the transmitted and reflected probability density profiles.

\end{abstract}

\maketitle


\section{Introduction}


Active particles are non-equilibrium entities that consume energy and convert it into directed motion~\citep{doi:10.1146/annurev-conmatphys-070909-104101}. They can be living organisms such as bacteria, algae, animals and birds, or inanimate entities such as colloids or robots~\citep{Gompper_2020}. They span a large range of scales from micrometer-sized molecular motors~\citep{Doostmohammadi2018} to crowds of humans covering hundreds of meters~\citep{annurev:/content/journals/10.1146/annurev-conmatphys-031620-100450}. A curious inanimate hydrodynamic active system is that of walking~\citep{Couder2005,Couder2005WalkingDroplets} and superwalking~\citep{superwalker} droplets. In this active system, periodically bouncing millimeter-sized droplets walk horizontally on the free surface of a vertically vibrating bath of the same liquid. Upon each impact, the droplet imprints a localized slowly decaying standing wave on the free surface. The droplet then interacts with these self-generated waves on subsequent impacts which guide the droplet motion and give rise to a self-propelled wave-particle entity (WPE). For high amplitudes of vertical bath vibrations, the droplet-generated waves decay very slowly in time, and the walking motion of the droplet is governed by the history of waves generated by the droplet along its trajectory, giving rise to path memory in this active system. Such active WPEs have been shown to mimic several peculiar features that are usually associated with the microscopic quantum realm~\citep{Bush2020review}.

Encouraged by the several hydrodynamic quantum analogs exhibited by these active WPEs, the framework of generalized pilot-wave dynamics~\citep{Bush2015} has been formulated which allows the exploration of a wider class of dynamical systems and parameter-space that are not restricted by experimental constraints. In particular, this theoretical framework has motivated exploration of idealized one-dimensional pilot-wave models to explore quantum analogs and non-equilibrium behaviors~\citep{phdthesismolacek,Durey2020lorenz,Durey2020,Valani2022ANM,ValaniUnsteady,Valanilorenz2022,Perks2023,Valani2024,Valani2024asym}. Since the motion of WPEs is driven by path memory, typical models take the form of integro-differential equations of motion ~\citep{Oza2013,Rahman2020review}. However, for $1$D active WPEs with certain simple wave forms, the infinite-dimensional system of integro-differential equation converts to a Lorenz-like nonlinear dynamical system comprising of a few ordinary differential equations (ODEs)~\citep{phdthesismolacek,Durey2020lorenz,ValaniUnsteady,Valanilorenz2022}. For these systems, the dynamics of the active WPE can be rationalized in terms of the nonlinear and chaotic dynamics of Lorenz-like systems~\citep{Durey2020lorenz,Valanilorenz2022,Valani2024asym}. Such reduced-order ODE models allow for rapid and detailed exploration of the parameter space and initial conditions of the system.

One of the quantum analogs exhibited by these active WPEs is that of tunneling~\citep{Eddi2009,tunnelingnachbin,tunneling2020,hubert2017,CARMIGNIANI2014237,Papatryfonos2022}. \citet{Eddi2009} first showed in experiments that when a walking droplet is incident on a submerged barrier, it can unpredictably cross the barrier. This was further explored in experiments by \citet{tunneling2020} who showed that the unpredictability is due to sensitivity in the vertical bouncing dynamics of the droplet when interacting with the barrier, as opposed to uncertainty in initial conditions or noise in the experiment. The above experiments were done in a regime where a single walker in free space moves steadily at a constant speed. Recent experiments have shown that at high memories, this steady walking state can become unstable and the walking droplet develops velocity oscillations~\citep{Bacot2019}. In this paper, we theoretically and numerically consider a one-dimensional setup of a WPE incident on a Gaussian potential barrier with a particular focus on the effects of unsteady walking dynamics arising in the high-memory regime. Our aim is to explore in detail the parameter space of the system as well as the space of initial conditions. A complete model that captures the vertical bouncing dynamics, horizontal walking dynamics as well as the wave evolution, would be computationally very expensive to perform such a detailed exploration. Hence, we resort to an idealized Lorenz-like ODE model to investigate in detail the dynamics and statistics of barrier crossing as a function of system parameters and initial conditions. Nevertheless, we expect that the interesting dynamical behaviors identified from the detailed exploration of our simplified model will guide in selecting parameter regimes to probe for future experimental work and numerical exploration with more accurate models. Furthermore, independent of the experimental system of walking droplets, our work demonstrates that a plethora of rich behaviors associated with barrier crossing can arise from the underlying nonlinear and chaotic dynamics of a simple pilot-wave model.

The paper is organized as follows. In Sec.~\ref{Sec: theory} we present the theoretical model and show that the system dynamics reduce to a perturbed Lorenz system. We then explore the steady states of the system and do a linear stability in Sec.~\ref{sec: lin stab full model}. We then investigate in detail, the dynamics of barrier crossing in Sec.~\ref{Sec: dynamics}, followed by the statistics of barrier crossing in Sec.~\ref{Sec: statistics}. We conclude in Sec.~\ref{Sec: conclusion}.

\section{Theoretical Model}\label{Sec: theory}

\begin{figure}
\centering
\includegraphics[width=\columnwidth]{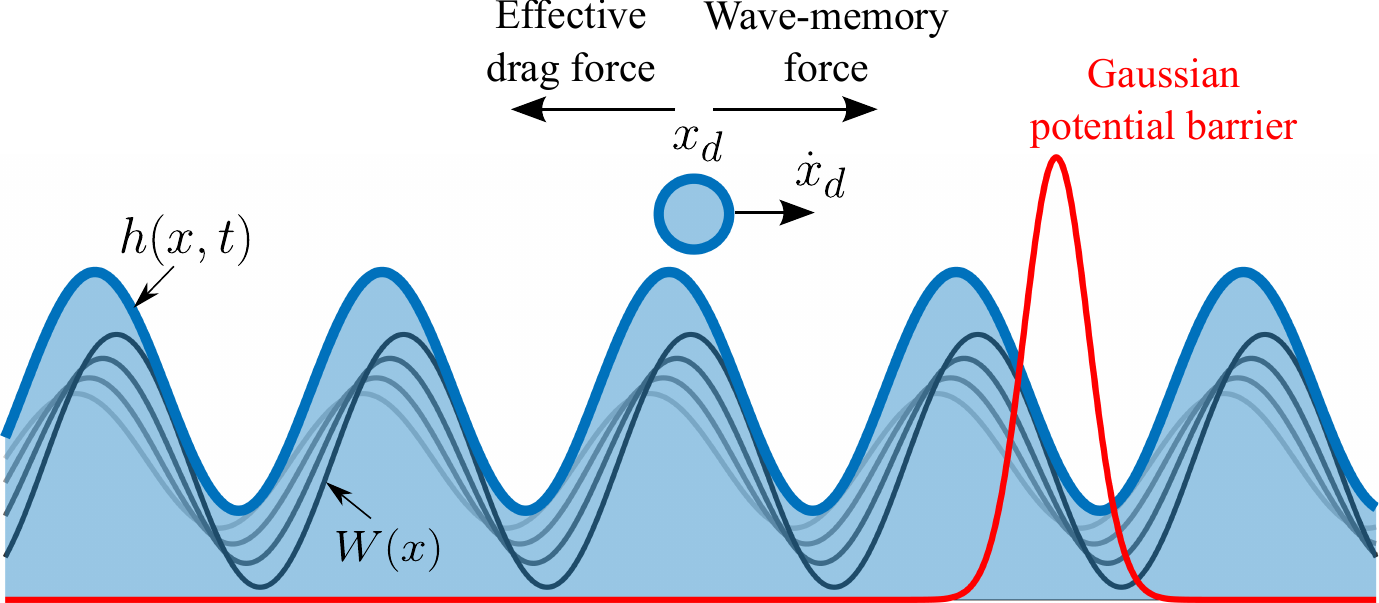}
\caption{Schematic of the system showing a one-dimensional active WPE (blue) directed at an isolated Gaussian potential barrier (red). The particle (blue circle) located at $x_d$ and moving horizontally with velocity $\dot{x}_d$ experiences -- a propulsion force from its self-generated wave field $h(x,t)$ (blue filled area), an effective drag force and an external force $F_{ex}$ when interacting with the Gaussian barrier. The underlying wave field  $h(x,t)$ is a superposition of the individual waves of spatial form $W(x)$ that decay exponentially in time (black and gray curves with the grayness indicating waves generated at earlier times), and are generated by the particle continuously along its trajectory.}
\label{Fig: schematic}
\end{figure}

As illustrated schematically in Fig.~\ref{Fig: schematic}, consider a particle (droplet) of mass $m$ at position $x_d$ and moving horizontally in one-dimension with velocity $\dot{x}_d$ towards an isolated Gaussian potential barrier $V(x)$. At each instance of time, the particle generates a standing wave with spatial form $ W(x)$ centered at the particle position and the wave decays exponentially in time. The horizontal walking motion of this one-dimensional WPE is governed by the following integro-differential equation of motion:~\citep{Oza2013,ValaniUnsteady} 
\begin{align}\label{Eq: dimensional eq}
    &m\ddot{x}_d+D\dot{x}_d = \\ \nonumber
    &\frac{F}{T_F}\int_{-\infty}^t f\left(k_F(x_d(t)-x_d(s))\right)\,\text{e}^{-(t-s)/T_F \text{Me}}\,\text{d}s\\ \nonumber
    &+\tilde{H}(x_d-\tilde{x}_b)e^{-\left(\frac{x_d-\tilde{x}_b}{\tilde{W}}\right)^2} . 
\end{align}
On the left side of Eq.~\eqref{Eq: dimensional eq}, there are two terms - the first representing the inertia of the particle and the second representing the effective drag force on the particle, where dots denote derivatives with respect to time. The first term on the right side of the equation represents the force exerted on the particle by its self-generated wave field $h(x,t)$. The wave field $h(x,t)$ is a superposition of the individual exponentially decaying waves of spatial form $W(x)$ that are generated by the particle continuously along its trajectory. This force is directly proportional to the gradient of the underlying wave field, where the function $f(x)=-W'(x)$ is the negative gradient of the individual wave form created by the particle. The second term on the right side is the external force arising from the Gaussian potential barrier $V(x)=V_0e^{-(x-\tilde{x}_b)^2/\sigma_0^2}$, and takes the following form:
\begin{align*}
    F_{ex}&=-\frac{dV}{dx}=\frac{2V_0}{\sigma_0}\left(x_d-\tilde{x}_b\right)\,\text{e}^{-\left(\frac{x_d-\tilde{x}_b}{\sigma_0}\right)^2}\\
		&= \tilde{H}\left(x_d-\tilde{x}_b\right)\,\text{e}^{-\left(\frac{x_d-\tilde{x}_b}{\tilde{W}}\right)^2}.
\end{align*}
Here, $\tilde{H}=2V_0/\sigma_0$ is the height to width ratio parameter of the Gaussian potential, and $\tilde{W} =\sigma_0 $ is the characteristic width of the Gaussian potential, where $V_0$ denotes the amplitude (height) of the Gaussian barrier at its center located at $\tilde{x}_b$. Other parameters in Eq.~\eqref{Eq: dimensional eq} are as follows: $D$ is an effective drag coefficient, $k_F=2\pi/\lambda_F$ is the Faraday wavenumber with $\lambda_F$ the Faraday wavelength, $F=m g A k_F$ is a non-negative wave-memory force coefficient where $g$ is the gravitational acceleration and $A$ is the amplitude of surface waves, $\text{Me}$ is the memory parameter that describes the proximity to the Faraday instability and $T_F$ is the Faraday period of droplet-generated standing waves and also the bouncing period of the walking droplet. We refer the interested reader to \citet{Oza2013} for more details and explicit expressions for these parameters.

Non-dimensionalizing Eq.~\eqref{Eq: dimensional eq} using $t'={Dt}/{m}$ and $x'=k_F x$, and dropping the primes on the dimensionless variables we get,
\begin{align}\label{Eq: dimless eq}
\ddot{x}_d+\dot{x}_d=&R\int_{-\infty}^t f(x_d(t)-x_d(s))\,\text{e}^{-(t-s)/\tau}\,\text{d}s\notag\\
    &+H(x_d-x_b)e^{-({x_d-x_b})^2/{W}^2}.
\end{align}
Here, $R=m^3gAk_F^2/D^3T_F$ and $\tau=DT_F\text{Me}/m$ denote the non-dimensional wave amplitude parameter and dimensionless memory parameter, respectively~\citep{Valani2024}. Moreover, $H={\tilde{H}m}/{D^2}$ is scaled height to width ratio parameter corresponding to the external Gaussian potential, $W=k_F\tilde{W}$ is a dimensionless Gaussian width, and $x_b=k_F\tilde{x}_b$ is a dimensionless position corresponding to the center of the Gaussian barrier.

For the one-dimensional WPE, we choose a simple sinusoidal particle-generated wave form such that $W(x)=\cos(x)$ and $f(x)=\sin(x)$, which allows us to transform the integro-differential equation of motion in \eqref{Eq: dimless eq} into a Lorenz-like system of ordinary differential equations (ODEs) giving us~\cite{phdthesismolacek,Durey2020lorenz,ValaniUnsteady,Valanilorenz2022} (see \citet{Valanilorenz2022} for a derivation)
\begin{equation}
    \label{lorenz}
    \begin{split}
    \dot{x}_d&=X \\
    \dot{X}&=Y-X+H(x_d-x_b)e^{-(x_d-x_b)^2/W^2}, \\
    \dot{Y}&=-\frac{1}{\tau} Y+XZ, \\
    \dot{Z}&=R-XY-\frac{1}{\tau} Z.
  \end{split}
\end{equation}

 These ODEs are rescaled Lorenz equations~\citep{Lorenz1963} with a couple of additions: (i) an added equation representing evolution of particle position, $\dot{x}_d=X$, and (ii) an added term arising from the external Gaussian potential barrier, $H(x_d-x_b)\,\text{e}^{-(x_d-x_b)^2/W^2}$, in the $\dot{X}$ equation in \eqref{lorenz}. Here, $X=\dot{x}_d$ is the particle's velocity, $$Y=R \int_{-\infty}^{t} \sin(x_d(t)-x_d(s))\,\text{e}^{-(t-s)/\tau}\,\text{d}s$$ is the wave-memory force, and $$Z=R\int_{-\infty}^{t} \cos(x_d(t)-x_d(s))\,\text{e}^{-(t-s)/\tau}\,\text{d}s$$ is the overall wave height at the particle location. 
In this paper, we numerically solve the system of ODEs described in Eq.~\eqref{lorenz} is simulated using MATLAB's built-in solver, ode45, with absolute and relative tolerance of $10^{-10}$. 


\section{Equilibrium states and their stability}\label{sec: lin stab full model}

First, we find the fixed points or equilibrium solutions of the dynamical system in Eq.~\eqref{lorenz}. Setting the time derivatives to zero in Eq.~\eqref{lorenz} results in the following equilibrium solution: 
\begin{equation*}
    x_d=x_b,\:X=0,\:Y=0\:,\:Z=\tau R.
\end{equation*}
This corresponds to a stationary state where the WPE remains stationary and located at the center of the applied Gaussian potential barrier. We note that for this WPE in free space (i.e. in the absence of the external Gaussian potential), there are two equilibrium solutions: (i) a translation invariant stationary state
\begin{equation*}
    x_d=x_0,\:X=0,\:Y=0\:,\:Z=\tau R,
\end{equation*}
where $x_0$ is any particle position, and (ii) a symmetric pair of steadily moving WPE at constant velocity~\citep{Valanilorenz2022,ValaniUnsteady}
\begin{equation*}
x_d=x_0+Xt,\:X=Y=\pm\sqrt{R-\frac{1}{\tau^2}}\:,\:Z=\frac{1}{\tau}.
\end{equation*}
In the context of the $(X,Y,Z)$ phase-space of the Lorenz system, the stationary state is a fixed point at the origin, while the left/right steady moving states are the symmetric fixed points located at the center of each wing of the Lorenz attractor.

We analyze the stability of the stationary state in the presence of the Gaussian potential barrier. To do this, we apply a small perturbation to the stationary state as follows~\citep{strogatz}: $(x_d,X,Y,Z)=(x_b,0,0,\tau R)+\epsilon(x_{d1},X_1,Y_1,Z_1)$, where $\epsilon>0$ is a small perturbation parameter. Substituting this in Eq.~\eqref{lorenz} and comparing terms of $O(\epsilon)$, we obtain the following linear system that governs the evolution of perturbations:
\begin{gather*}
 \begin{bmatrix} 
 \dot{x}_{d1} \\
 \dot{X}_1 \\
 \dot{Y}_1 \\
 \dot{Z}_1 
 \end{bmatrix}
 =
  \begin{bmatrix}
0 & 1 & 0 & 0 \\
H & -1 & 1 & 0 \\
0 & R\tau & -\frac{1}{\tau} & 0 \\
0 & 0 & 0 & -\frac{1}{\tau}
 \end{bmatrix}
  \begin{bmatrix}
  x_{d1}\\
  {X}_1 \\
 {Y}_1 \\
 {Z}_1 
 \end{bmatrix}.
\end{gather*}
The linear stability is determined by the eigenvalues of the right-hand-side matrix. This results in the following characteristic polynomial equation to be solved for the eigenvalues $\lambda$ which determines the growth rate of perturbations:
\begin{align}\label{eq: polynomial}
\left(\lambda\tau+1\right) \left(\tau\lambda^3+\left(\tau+1\right)\lambda^2+\left(1-R\tau^2-H\tau\right)\lambda-H\right)=0.
\end{align}
We can obtain non-trivial eigenvalues by solving the cubic equation of the characteristic polynomial in Eq.~\eqref{eq: polynomial}. However, by invoking Descartes' rule of sign, one can deduce information about the nature of the real eigenvalues of this cubic equation without explicitly solving the cubic equation. 	The number of sign changes between consecutive non-zero coefficients is one as per below:
	\begin{align*}
	\tau\lambda^3&&+\left(\tau+1\right)\lambda^2&&+\left({1}-R\tau^2-H\tau\right)\lambda&&-H=0.\\
		(+)  &&(+)&&(+/-)&&(-)
	\end{align*}
So, we deduce the existence of atleast one real positive root for Eq.~\eqref{eq: polynomial} and hence the stationary state at the center of the Gaussian barrier is unstable.

Now the stability of the stationary state without the external force, i.e. $ H=0 $, is governed by the characteristic equation:
\begin{align*}
		\tau\lambda^3+\left(\tau+1\right)\lambda^2+\left(1-R\tau^2\right)\lambda&=0,
	\end{align*}
which gives us the following quadratic equation to solve for non-trivial eigenvalues:
	\[\lambda^2+\left(1+\frac{1}{\tau}\right)\lambda+\frac{1}{\tau}-R\tau=0,\]
 with the discriminant
	\[\Delta=\left(1+\frac{1}{\tau}\right)^2-4\left(\frac{1}{\tau}-R\tau\right)=\left(1-\frac{1}{\tau}\right)^2+4R\tau>0\]
	and eigenvalues
	\[\lambda_1 = \frac{-\left(1+\frac{1}{\tau}\right)+\sqrt{\Delta}}{2}, \quad \lambda_2 = \frac{-\left(1+\frac{1}{\tau}\right)-\sqrt{\Delta}}{2}<0.\]
Since $\lambda_2$ is always negative, the stationary state without the external potential barrier is stable when $\lambda_1<0$, and unstable when $\lambda_1>0$. This gives us the following equation that separates the stable and unstable stationary state solutions without the external potential in $(\tau,R)$ parameter space\\
	\begin{equation}\label{eq: stability SS no barrier}
	 R = \frac{1}{\tau^2}.
	\end{equation}
Above this curve once the stationary state is unstable, one gets steady walking solutions and then eventually unsteady walking solutions. The regime without the external potential is valid when the WPE is sufficiently far away from the center of the Gaussian potential barrier i.e. when $|x_d(t)-x_b|\gg W$, since the Gaussian profile decays quickly away from its peak. Hence, when the WPE is not interacting with the Gaussian barrier, its dynamics are governed by the dynamics of the Lorenz system. In other words, our system can be thought of as a point evolving in the phase-space of the Lorenz system which is perturbed when the WPE comes close to the potential barrier.

\begin{figure}
\centering
\includegraphics[width=\columnwidth]{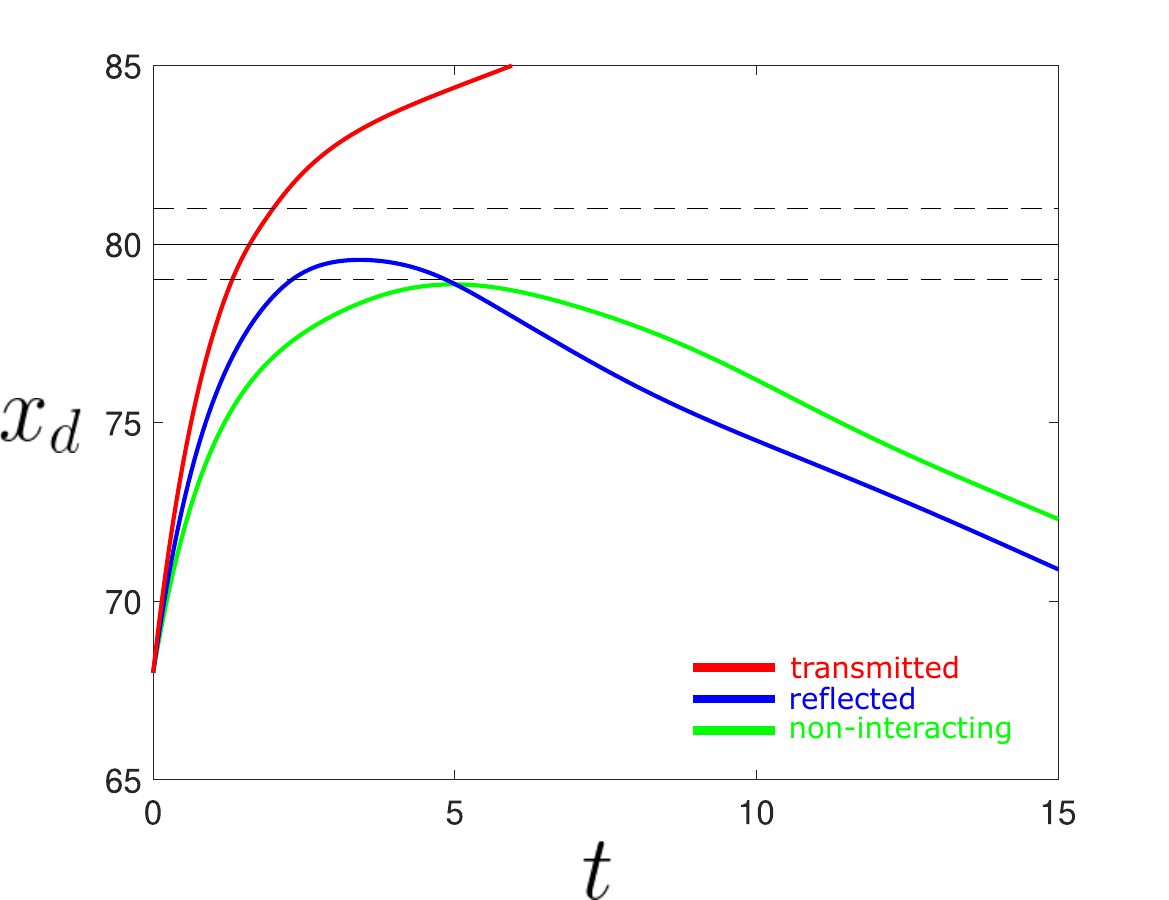}

\caption{ Classification of WPE trajectories when interacting with a Gaussian potential barrier. We classify the WPE motion into three types: (red) WPE transmitted through the potential barrier, (blue) WPE reflecting from the potential barrier, and (green) WPE not (strongly) interacting with the potential barrier. The Gaussian potential barrier is centered at $ x_{b} = 80$ (black solid line) with dimensionless width $W = 1$ (black dashed line represent $x_b\pm W$). For the trajectories shown, the WPE started at $x_d(0)=68$ with initial velocities $X_0 = 15$ (red), $12$ (blue) and $10$ (green). Other parameters were fixed to $R=1$, $\tau=1.5$, $H=2.5$, $Y(0)=0$ and $Z(0)=0$.
}
\label{Fig:trajectory types}
\end{figure}

\section{Dynamics of barrier crossing}\label{Sec: dynamics}

\subsection{Trajectory classification}\label{Sec: trajectory classification}
We now turn to explore the dynamical features of the WPE when interacting with the potential barrier. Since we are primarily interested in the long-time behavior of the WPE, we classify the interaction of the WPE with the potential barrier into three types as shown in Fig.~\ref{Fig:trajectory types}: (i) transmitted trajectories (red) - this is when the WPE crosses the potential barrier and stays on that side at long times, (ii) reflected trajectories (blue) - this is when the WPE interacts with the potential barrier but it is found on the same side as the incident trajectory at long times, and (iii) non-interacting trajectories (green) - this is when the WPE does not strongly interact with the potential barrier i.e. when $|x_d(t)-x_b|> W$ for all times. We note that for the transmitted and reflected trajectories, our classification includes the cases when there might be multiple interactions with the potential barrier, since our classification is based on the side of the barrier that the WPE finds itself at the end of the simulation. Based on the WPE trajectory classification into these three states, we now turn to explore the effects of system parameters and initial conditions on which of the three states are realized for the system.

\subsection{Dynamics of barrier crossing with increasing memory}\label{Sec: dynamics with memory}

Since the WPE dynamics without the potential barrier maps to the Lorenz system~\citep{ValaniUnsteady,Valani2024}, we proceed by reviewing different attractors realized in the Lorenz model~\citep{Sparrowbook,Valani2024} of Eq.~\eqref{lorenz} (with $H=0$), as a function of the memory parameter $\tau$. For a fixed $R$ and small values of $\tau$ (i.e. $\tau<1/\sqrt{R}$), there is only one equilibrium point of the system at $(X,Y,Z)=(0,0,0)$ which is stable. This corresponds to a stationary non-walking state for the particle since its velocity is $\dot{x}_d=X=0$. With increasing $\tau$, the Lorenz system undergoes a pitchfork bifurcation (at $\tau=1/\sqrt{R}$) where the non-walking state becomes unstable and a symmetric pair of stable equilibrium points emerge, $(X,Y,Z)=(\pm\sqrt{R-1/\tau^2},\pm\sqrt{R-1/\tau^2},1/\tau)$; this corresponds to steady walking motion of the particle to right/left. Further increase in $\tau$ leads to a complex set of global bifurcations in the Lorenz system~\citep{Sparrowbook,jackson_1990} that lead to oscillatory dynamics of $X$ and eventually chaos emerges at large $\tau$ with a strange attractor in the phase space. Phase-space dynamics on the Lorenz chaotic attractor corresponds to a chaotic walk~\citep{ValaniUnsteady,Valaniattractormatter2023} for the active particle where it unpredictably switches between left and right moving states as the phase space trajectory jumps between left and right ``wing" of the Lorenz attractor. The stationary 
non-walking state as well as the steady walking states of the WPE are routinely observed in experiments with walking and superwalking droplets moving in free space~\citep{Couder2005WalkingDroplets,superwalker}. Instability of the steady walking state and the presence of velocity oscillations has also been reported for a free walker~\citep{Bacot2019}, whereas chaotic walking states at high memory of a WPE moving in free-space have not been realized experimentally yet.

\begin{figure*}
\centering
\includegraphics[width=2\columnwidth]{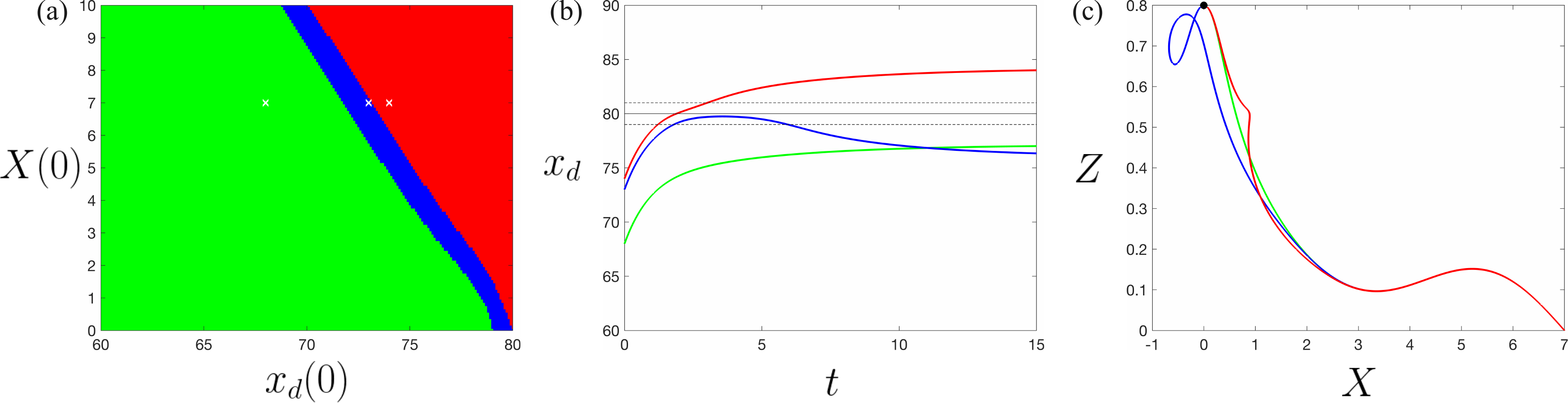}
\caption{Crossing dynamics in the stationary state regime of the WPE. (a) Basin of attraction plot showing initial conditions color-coded based on the three different types of WPE trajectories: transmitted (red), reflected (blue) and non-interacting (green), as classified in Sec.~\ref{Sec: trajectory classification}. For each type of trajectory, a typical space-time trajectory is shown in (b) and the corresponding $(X,Z)$ phase-plane projection is shown in (c). The initial conditions corresponding to trajectories in (b) and (c) are marked by a white $\times$ in (a) and they correspond to $(x_d(0),X(0))=(68,7)$ (green), $(73,7)$ (blue) and $(74,7)$ (red). The Gaussian potential barrier as shown in (b) is centered at $ x_{b} = 80$ (black solid line) with width $ W = 1 $ (black dashed line represent $x_b\pm W$). The black filled circle in (c) corresponds to the stationary point of the free WPE $(X,Y,Z)=(0,0, R\tau)$. Other parameter values are fixed to $R=1$, $H=1.5$, $\tau=0.8$, $Y(0)=0$, $Z(0)=0$ and simulations were run till $t_{end}=500$. See also Movie 1.}
    \label{Fig:Static State}
\end{figure*}

We proceed by exploring in detail the regime of small memory $\tau$ where the Lorenz system has a stable fixed point at the origin and hence the WPE has a non-walking equilibrium steady state. However, the introduction of the potential barrier makes our dynamical system deviate from the Lorenz system and we find different outcomes, as classified in Sec.~\ref{Sec: trajectory classification}, based on initial conditions. We fix $R=1$ and $\tau=0.8$, and since $R < \frac{1}{\tau^2}$, we are in the stationary state regime. In Fig.~\ref{Fig:Static State}(a), we color the initial condition plane $(x_d(0),X(0))$ based on our three types of trajectories and an example of each type of trajectory is shown in Figs.~\ref{Fig:Static State}(b) and (c)~(see also Movie 1). We find that if the WPE starts far away from the Gaussian barrier and/or with a lower initial positive velocity, we get the green non-interacting trajectory since the WPE asymptotes towards the stationary state before getting near the barrier. However, if the WPE starts near the barrier and/or has a high initial positive velocity, then the effects of initial conditions allow the WPE to reach the potential barrier, and it can transmit or reflect from the barrier before reaching its stationary non-walking state. Hence, the transmitted and reflected trajectories in this regime are purely due to the transient effects of initial conditions since the steady state is a stationary non-walking state for the WPE.

\begin{figure*}
\centering
\includegraphics[width=2\columnwidth]{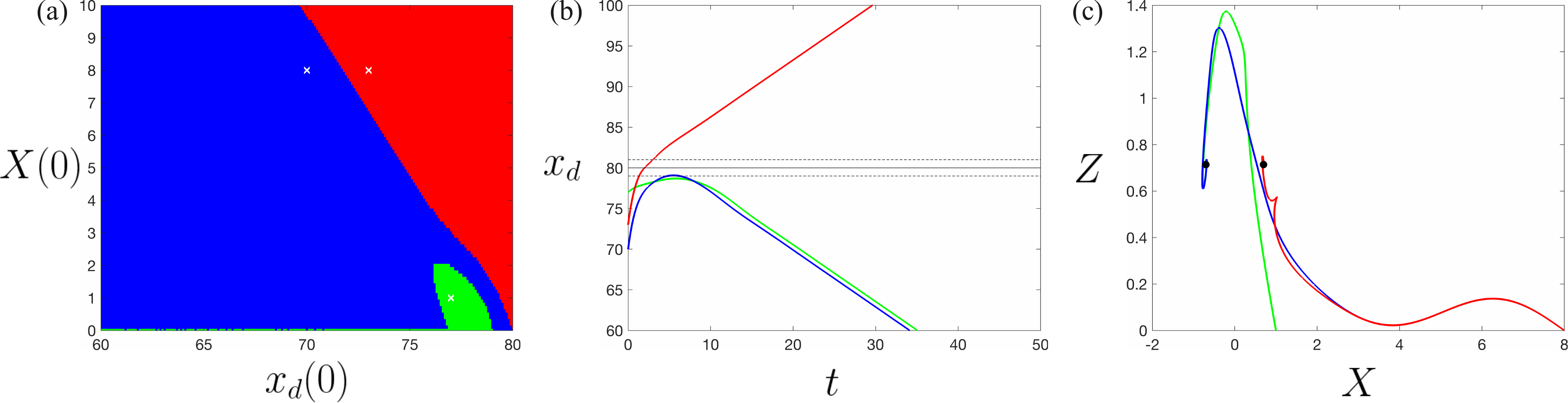}
\caption{Crossing dynamics in the steady walking regime of the WPE. Basin of attraction plot showing initial conditions color-coded based on the three different types of WPE trajectories: transmitted (red), reflected (blue) and non-interacting (green), as classified in Sec.~\ref{Sec: trajectory classification}. For each type of trajectory, a typical space-time trajectory is shown in (b) and the corresponding $(X,Z)$ phase-plane projection is shown in (c). The initial conditions corresponding to trajectories in (b) and (c) are marked by a white $\times$ in (a) and they correspond to $(x_d(0),X(0))=(77,1)$ (green), $(70,8)$ (blue) and $(73,8)$ (red). The Gaussian potential barrier as shown in (b) is centered at $ x_{b} = 80$ (black solid line) with width $ W = 1 $ (black dashed line represent $x_b\pm W$). The black filled circles in (c) correspond to the steady walking equilibrium points of the free WPE $(X,Y,Z)=(\pm\sqrt{R-1/\tau^2}, \pm\sqrt{R-1/\tau^2}, 1/\tau )$. Other parameter values are fixed to $R=1$, $H=1.5$, $\tau=1.4$, $Y(0)=0$, $Z(0)=0$ and simulations were run till $t_{end}=500$. See also Movie 2.}
\label{Fig:Steady Walking State}
\end{figure*}

We now proceed to explore the barrier crossing dynamics just above the pitchfork bifurcation when the WPE has stable steady walking states. Figure~\ref{Fig:Steady Walking State} show the same set of plots as Fig.~\ref{Fig:Static State} but now for a slightly larger value of $\tau=1.4$~(see also Movie 2). The phase-space trajectory in the Lorenz system without the potential barrier converges to the fixed points $(X,Y,Z)=(\pm\sqrt{R-{1}/{\tau^2}},\pm\sqrt{R-{1}/{\tau^2}},{1}/{\tau})$ which corresponds to a steadily walking WPE. In Fig.~\ref{Fig:Steady Walking State}(a) we see that if the WPE starts further away from the barrier we get a blue reflected trajectory. This is because the WPE motion settles to its steady velocity $X= \sqrt{R-{1}/{\tau^2}}$ before reaching the barrier, and this velocity is not enough to cross the potential barrier; hence the WPE is reflected from the barrier. We find red transmitted trajectories for WPE starting near the barrier and/or with larger initial velocities. Again, due to the transient effects of larger initial positive velocities, WPE can cross the potential barrier that it would otherwise not be able to cross with its steady velocity. We also get a small green region for a WPE starting near the barrier with low initial velocities. The WPE starting with these set of initial conditions is unable to get very close to the potential barrier before it gets reflected, and hence it gets classified as green non-interacting trajectory. An example trajectory for each type of behavior as a space-time plot and a phase-plane projection is shown in Fig.~\ref{Fig:Steady Walking State}(b) and (c), respectively. Since the basins of attraction for transmitted and reflected states are smoothly partitioned in the initial condition space (see Fig.~\ref{Fig:Steady Walking State}(a)), in this regime we would not expect sensitivity in crossing the potential barrier based on initial conditions.

\begin{figure*}
\centering
\includegraphics[width=2\columnwidth]{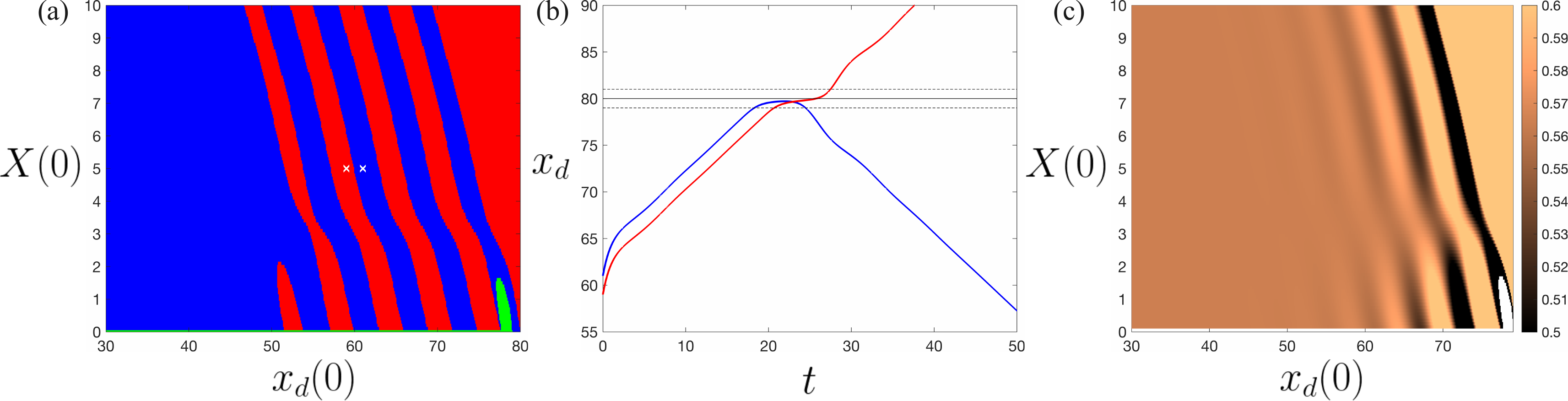}
\caption{Crossing dynamics in the transient velocity oscillations regime of the WPE. Basin of attraction plot showing initial conditions color-coded based on the three different types of WPE trajectories: transmitted (red), reflected (blue) and non-interacting (green), as classified in Sec.~\ref{Sec: trajectory classification}. For transmitted and reflected trajectories, a typical space-time evolution is shown in (b). The initial conditions corresponding to trajectories in (b) are marked by a white $\times$ in (a) and they correspond to $(61,5)$ (blue) and $(59,5)$ (red). Panel (c) shows a velocity colormap in the initial condition space of WPE velocity when it first encounters the barrier i.e. $x_d(t)=x_b-W$. The Gaussian potential barrier as shown in (b) is centered at $ x_{b} = 80$ (black solid line) with width $ W = 1 $ (black dashed line represent $x_b\pm W$). Other parameter values are fixed to $R=1$, $H=1.5$, $\tau=1.81$, $Y(0)=0$, $Z(0)=0$ and simulations were run till $t_{end}=500$. See also Movie 3.}
    \label{Fig:Steady Walking State2}
\end{figure*}

Further increasing memory parameter to $\tau=1.81$, the steady walking equilibrium points are still stable, however, their eigenvalues have become complex conjugates and they acquire the character of stable spirals in phase space. Hence, the approach to these fixed points in phase-space will have an oscillatory component and this corresponds to velocity oscillations of the WPE as it reaches its steady state velocity. Figure~\ref{Fig:Steady Walking State2}(a) shows the basin of attraction of different types of trajectories in the initial condition plane. For WPE starting far away from the barrier, the transient effects have decayed and the WPE converges to the steady walking state when it approaches the barrier. Since this steady walking speed is still not sufficient to cross the potential barrier, we have reflected trajectories and a corresponding blue region for $x_d(0)<<x_b$. For an intermediate range of starting positions away from the potential barrier, we see alternating red and blue stripes in Fig.~\ref{Fig:Steady Walking State2}(a). These patterns in the basin of attraction arise from velocity oscillations in the WPE motion as it approaches the barrier. Since the transient velocity oscillations have not decayed by the time the WPE reaches the barrier, the WPE can have a larger or a smaller velocity than the critical velocity required to cross the barrier when it starts interacting with the barrier~(see Fig.~\ref{Fig:Steady Walking State2}(b)-(c)) and this dictates whether the WPE transmits through the potential barrier or gets reflected from it~(see also Movie 3). Again, due to similar mechanisms as shown for Fig.~\ref{Fig:Steady Walking State}, we have a small green region of non-interacting trajectories and red triangular region of transmitted trajectories for large initial positive velocities near the barrier. 

\begin{figure*}
\centering
\includegraphics[width=2\columnwidth]{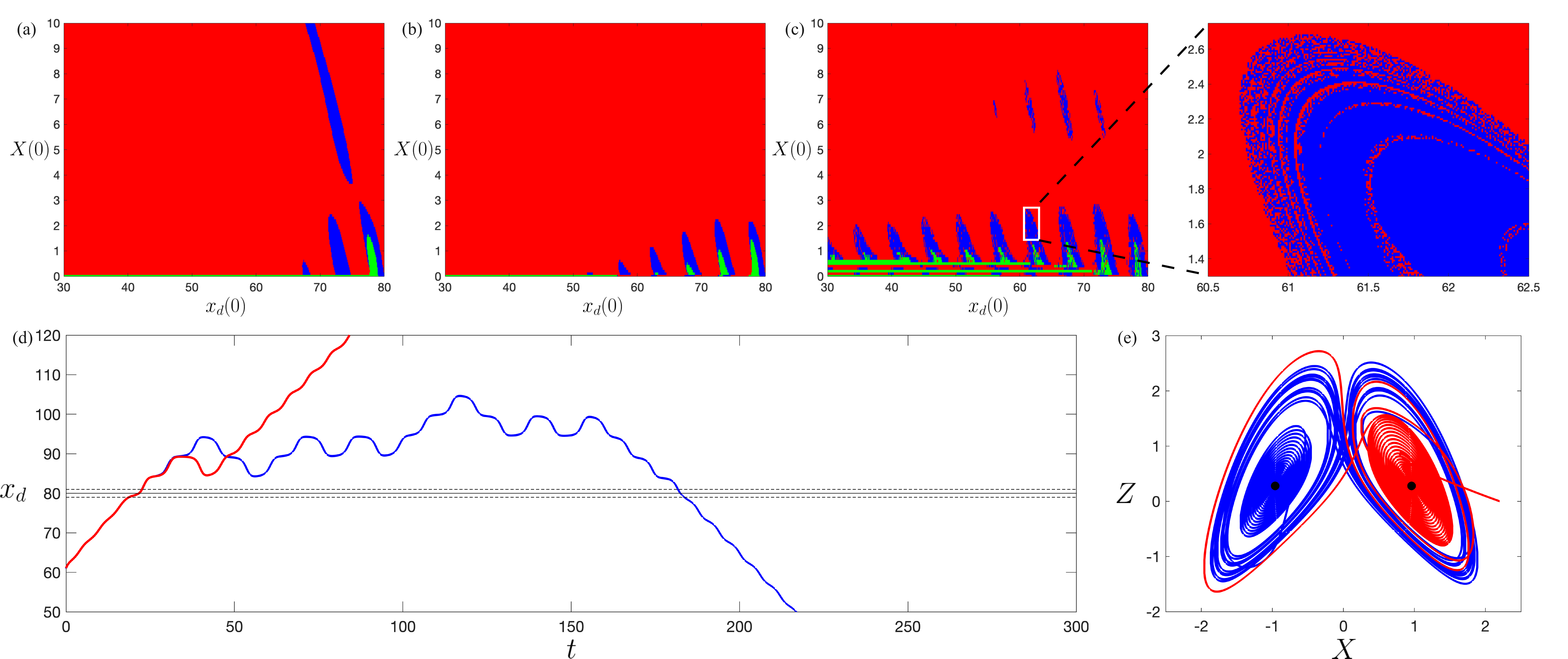}
\caption{Crossing dynamics in the transient chaotic regime of the WPE. Basin of attraction plot showing initial conditions color-coded based on the three different types of WPE trajectories: transmitted (red), reflected (blue) and non-interacting (green), as classified in Sec.~\ref{Sec: trajectory classification}, for (a) $\tau=2$, (b) $\tau=3$ and (c) $\tau=3.6$. The rightmost panel shows a blow-up of panel (c). A typical (d) space-time trajectory and (e) phase-plane trajectory for two closely spaced initial conditions $(x_d(0),X(0))=(61.07,2.2)$ (red) and $(x_d(0),X(0))=(61,2.2)$ (blue) corresponding to the blow-up region in panel (c) where transient chaos is realized. The Gaussian potential barrier as shown in (b) is centered at $ x_{b} = 80$ (black solid line) with width $ W = 1 $ (black dashed line represent $x_b\pm W$). Other parameter values are fixed to $R=1$, $H=1.5$, $Y(0)=0$, $Z(0)=0$ and simulations were run till $t_{end}=500$ except for the blow-up region where the simulations were run until $t_{end}=1000$. See also Movie 4.}
    \label{Fig:Steady Walking State3}
\end{figure*}

As the memory parameter is increased to $\tau=2$ and $\tau=3$, then as shown in Figs.~\ref{Fig:Steady Walking State3}(a)-(b), we see two qualitative changes. Firstly, the steady state velocity of the WPE is now sufficient to cross the potential barrier and hence, for WPE starting further away from the barrier we now see transmitted trajectories and a corresponding red region. Secondly, the extent of blue and red stripes are greatly reduced since even with velocity oscillations, the minimum velocity is still enough to cross the potential barrier for most part of the initial condition space shown in Figs.~\ref{Fig:Steady Walking State3}(a)-(b). As the memory parameter is further increased to $\tau=3.6$, as shown in Fig.~\ref{Fig:Steady Walking State3}(c), we see emergent patches of intermingled basins~(see the inset of Fig.~\ref{Fig:Steady Walking State3}(c)) between the transmitted and reflected states due to the presence of transient chaos in the Lorenz system in this parameter range. This is illustrated by showing the position-time~(see Fig.~\ref{Fig:Steady Walking State3}(d)) and projection of phase-space~(see Fig.~\ref{Fig:Steady Walking State3}(e)) of two trajectories with closely spaced initial conditions in the intermingled basin region. These two closely starting WPEs follow almost identical trajectories in phase-space before interacting with the potential barrier. As they both interact and cross the potential barrier, their almost identical phase-space trajectories are perturbed in different ways. One of the perturbed phase-space trajectories (red), quickly settles back on the same side of the spiral fixed point after a short excursion on the other wing of the Lorenz attractor. The other perturbed phase-space trajectory (blue) gets trapped in transient chaos after crossing the potential barrier. This transient chaos results in chaotic walking for this WPE and it interacts with the potential barrier again and crosses it. During this second interaction, the phase-space trajectory is again perturbed and settles towards the negative velocity spiral fixed point and the WPE ends up on the same side of the potential barrier as the incident trajectory~(see also Movie 4). Hence we see a final state sensitivity~\citep{GREBOGI1983415} between transmitted and reflected states due to the presence of transient chaos at high memories for the WPE. 

For even larger memory $\tau$, which corresponds to fully developed chaos in the Lorenz system, the WPE performs a chaotic walk for all times and crosses the potential barrier multiple times. Hence, the WPE can be found on either side of the barrier at the end of the simulations and the side on which it is found will vary with the simulation time. Further, the basin of attraction for transmitted and reflected states become completely intermingled. Since there is no coherent/steady motion of the WPE in this regime even far away from the potential barrier, it is not the most interesting regime in the setup of our WPE crossing the potential barrier.

\section{Statistics of barrier crossing}\label{Sec: statistics}

After having explored the different dynamical mechanisms for barrier crossing in the previous section, we now turn to investigate the statistical properties for barrier crossing. We do this by calculating the transmission probability $\text{Pr}(T)$. This probability is calculated by ensemble averaging over a subset of possible initial conditions. Specifically, we consider the plane of initial conditions where $x_d(0)\in[30,70]$ and $X(0)\in[0,3\sqrt{R-1/\tau^2}]$ with fixed $Y(0)=Z(0)=0$. This subset of initial conditions corresponds to no initial wave field for the WPE and the initial location of the WPE is far from the Gaussian potential centered at $x_b=80$. Further, the WPE initial velocity is in a reasonable range of its steady velocity. These choice of initial conditions would be reasonable for the experimental walking droplet system, and they also ensure that the crossing statistics are not significantly distorted by rapidly decaying transient effects from initial conditions. The transmission probability is then defined as the fraction of area occupied by red region in the initial condition space relative to the total area covered by blue and red regions (i.e. we discard the green region).

\subsection{Transmission probability as a function of memory}\label{Sec: TP memory}

\begin{figure}
\centering
\includegraphics[width=0.9\columnwidth]{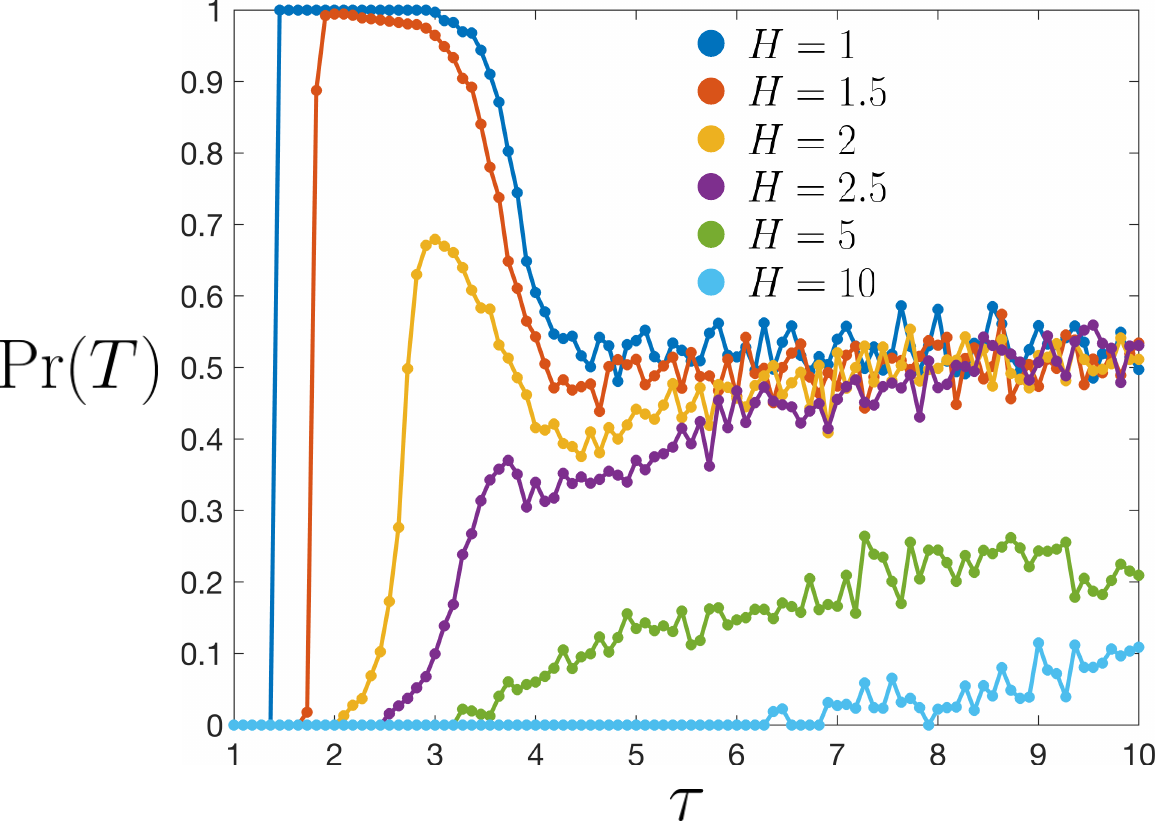}
\caption{Crossing statistics for the WPE as a function of the memory parameter $\tau$ for different values of the dimensionless height $H$ of the potential barrier. Plot of probability of transmission across the potential barrier as a function of $\tau$ for $H=1, 1.5, 2, 2.5, 5$ and $10$. The probability was calculated by uniformly sampling $400$ different points in the initial condition space $x_d(0)\in[30,70]$ and $X(0)\in[0,3\sqrt{R-1/\tau^2}]$. The Gaussian potential barrier is centered at $ x_{b} = 80$ with width $W = 1$. Other parameter values are fixed to $R=1$, $Y(0)=0$, $Z(0)=0$ and simulations were run till $t_{end}=500$.}
    \label{Fig:Prob crossing}
\end{figure}

In Fig.~\ref{Fig:Prob crossing}, we plot the transmission probability of the WPE as a function of the memory parameter $\tau$, for different values of the dimensionless height to width ratio $H$ and a fixed dimensionless width $W=1$. Since we have fixed $W=1$, a variation in $H$ can also be thought of as the variation in the height of the potential barrier when all other WPE intrinsic parameters are fixed. Hence, we will refer to $H$ as a dimensionless height when $W$ is fixed. For a fixed $R=1$ and $\tau<1$, the equilibrium states of the WPE are stationary non-walking states and hence we would not have any transmission for the set of initial conditions chosen~(see Fig.~\ref{Fig:Static State}). For $\tau>1$, transmission is possible and will depend on the parameter $H$. For $H=1$ and $H=1.5$, the steady walking velocity of the WPE just above $\tau=1$ is not sufficient to cross the barrier, and hence one still gets a zero probability of transmission just above $\tau=1$ for the set of initial conditions chosen~(see Fig.~\ref{Fig:Steady Walking State}). Once $\tau$ is large enough so that the steady WPE has enough velocity to cross the barrier, we see a sudden jump in the transmission probability from zero to one. Such a discontinuous jump would be expected for a classical particle crossing a potential barrier, since it cannot cross the potential barrier for low energies and it can cross the potential barrier for high energies with no possibility of partial crossing. For a quantum particle, the transmission probability varies smoothly as a function of the incident particle's energy~\citep{quantumgaussian}. In our system, after the discontinuous jump, the probability starts smoothly dropping for larger $\tau$ due to the dynamical effects of velocity oscillations and transient chaos as discussed in Figs.~\ref{Fig:Steady Walking State2}-\ref{Fig:Steady Walking State3}. Once $\tau$ is large enough that the WPE trajectories are chaotic, the transmission probability fluctuates near $0.5$. This is because in this regime, the barrier does not significantly influence the WPE dynamics. The WPE can cross the barrier multiple times and its final state (red or blue) will be equally likely at the end of the simulation time. 

For a larger value of $H=2$ and $2.5$, a larger value of memory $\tau$ is required before transmission is possible. Hence, when the WPE is first able to cross the barrier, it has already developed velocity oscillations and/or transient chaos. So the transmission probability initially varies smoothly with $\tau$ but it never reaches one. Once chaos fully develops, the transmission probability starts fluctuating and approaches a value of $0.5$. For even higher values of $H=5$ and $H=10$, the WPE is only able to cross the potential barrier for high enough $\tau$ where it is already in the chaotic regime. Hence, one sees fluctuations in the probability of transmission as soon as they become non-zero. Further, the probability of transmission does not readily approach the value $0.5$ since for large $H$, each interaction of the chaotic WPE with the potential barrier does not always result in a transmission event. 

We note that these absolute values of transmission probability depend on the choice of initial conditions, since the area occupied by different final states in the basin of attraction varies with the range of initial conditions chosen.


\subsection{Transmission probability as a function of barrier height}\label{Sec: TP height}

\begin{figure}
\centering
\includegraphics[width=0.9\columnwidth]{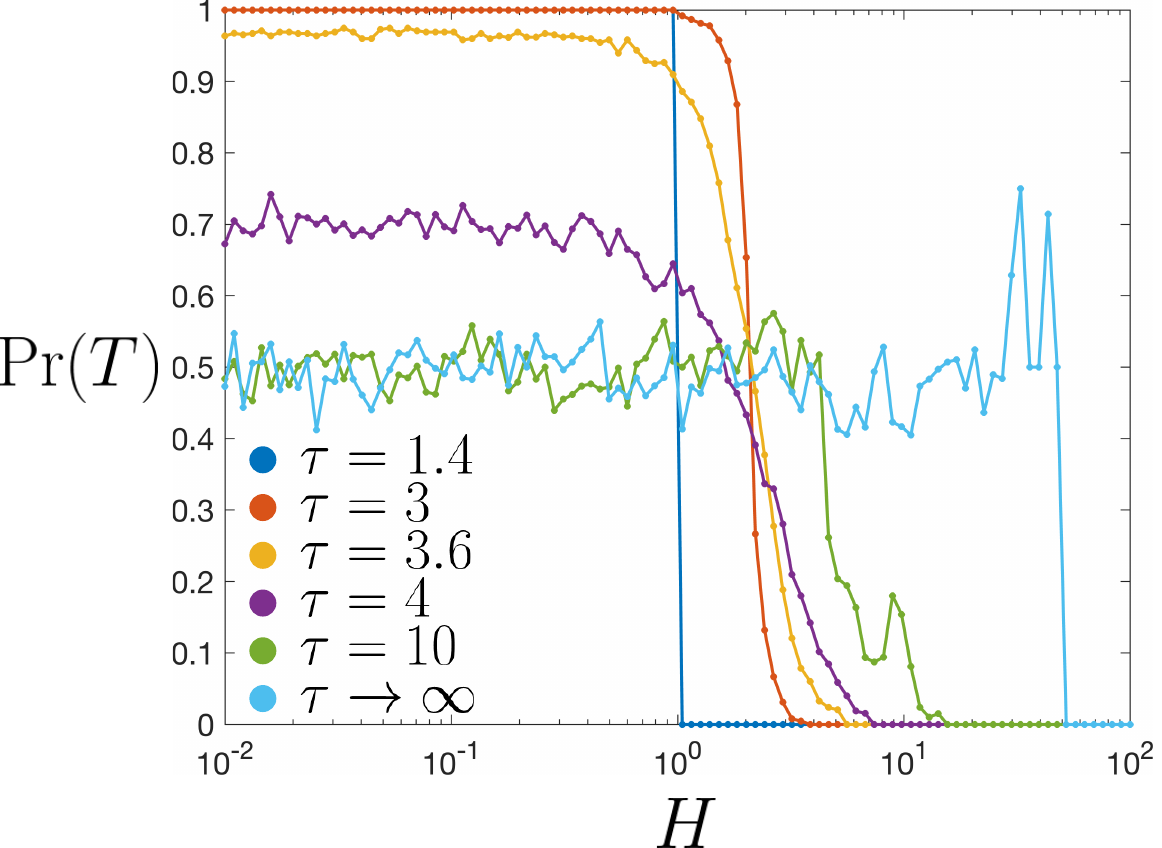}
\caption{Crossing statistics for the WPE as a function of the dimensionless barrier height $H$ for different values of the memory parameter $\tau$. Plot of probability of transmission across the potential barrier as a function of $H$ for $\tau=1.4, 3, 3.6, 4, 10$ and $\tau \xrightarrow{} \infty$. The probability was calculated by uniformly sampling $400$ different points in the initial condition space $x_d(0)\in[30,70]$ and $X(0)\in[0,3\sqrt{R-1/\tau^2}]$. The Gaussian potential barrier as shown in (b) is centered at $ x_{b} = 80$ (black solid) with width $ W = 1 $ (black dashed line represent $x_b\pm W$). Other parameter values are fixed to $R=1$, $Y(0)=0$, $Z(0)=0$ and simulations were run till $t_{end}=500$.}
\label{Fig:Prob crossing inf mem}
\end{figure}

We now turn to explore the transmission probability as a function of $H$ for different values of the memory parameter $\tau$. This is shown in Fig.~\ref{Fig:Prob crossing inf mem}. For a small value of memory $\tau=1.4$ just above the walking threshold, we find that the WPE always crosses the barrier for dimensionless height $H\lesssim1$ whereas barrier above this height are not crossed due to insufficient velocity of the WPE. This maximum height that the WPE can cross increases with increasing memory $\tau$. Further, the jump in transmission probability from one to zero get smoother as the WPE develops velocity oscillations and transient chaos for higher memory values of $\tau=3$ and $3.6$. For an even larger memory of $\tau=4$ and $\tau=10$ which corresponds to a chaotic WPE, we find that the WPE crosses even the smaller height barriers with a probability less than one and this probability then further decays to zero with increasing height. Even in the limit $\tau \xrightarrow{} \infty$, the maximum barrier height that the WPE can cross is finite. This shows that the WPE cannot indefinitely build up its wave field at infinite memory and cross any height $H$ of the potential barrier. The chaotic motion of the WPE at infinite memory results in both constructive and destructive interference of the underlying waves, and thus limiting the coherent build-up of the wave field~\citep{Valani2024}. However, the largest height $H$ of the barrier that WPE can cross at infinite memory will change with the dimensionless wave-amplitude parameter $R$.


\begin{figure}
\centering
\includegraphics[width=0.9\columnwidth]{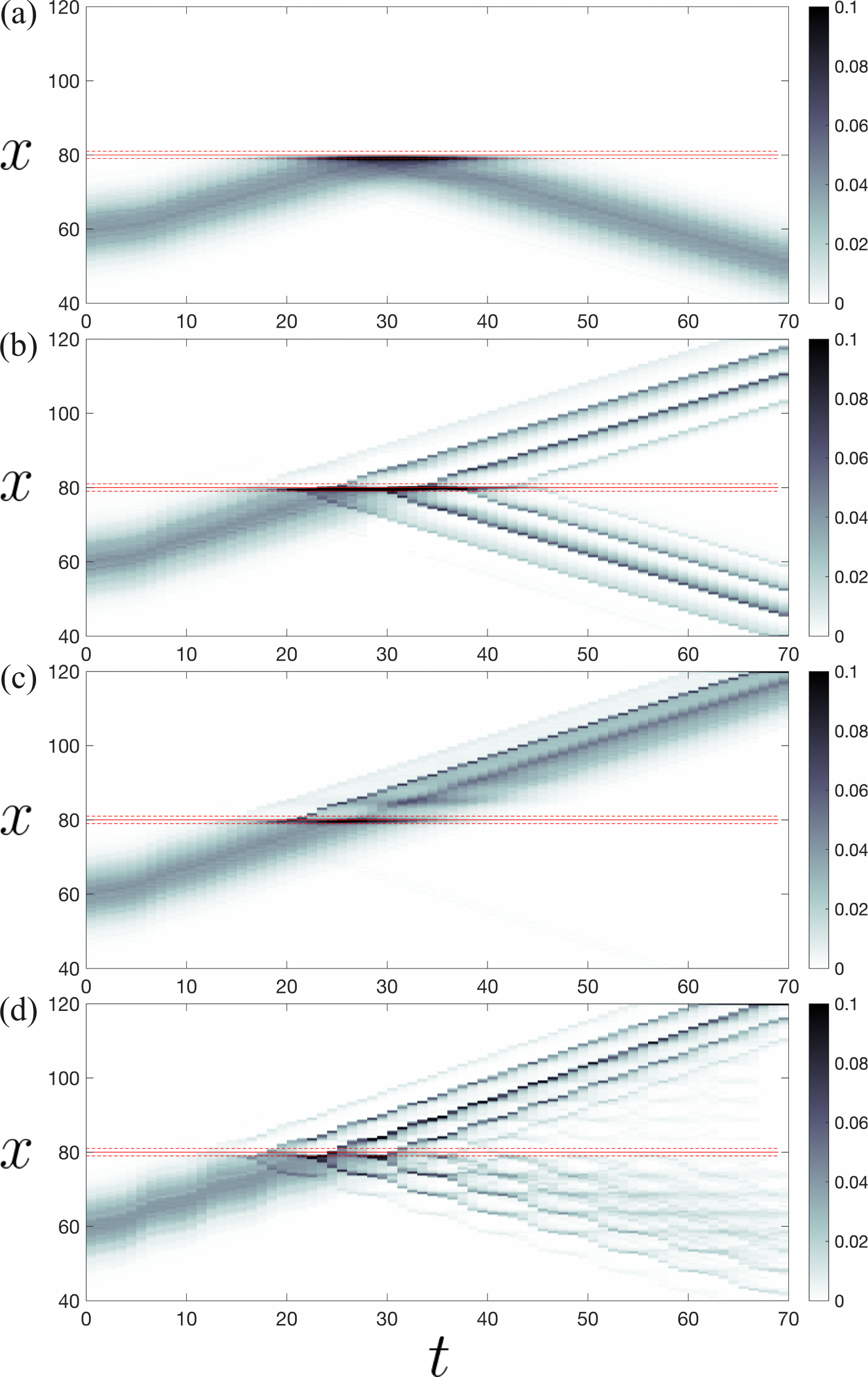}
\caption{Evolution of probability density of WPE's position (grayscale) as it interacts with a Gaussian potential barrier for memory parameters (a) $\tau=1.5$, (b) $\tau=1.81$, (c) $\tau=2$ and (d) $\tau=3.6$. The plot shows probability density from $10000$ different initial conditions with initial position chosen from a normal distribution with mean $60$ and standard deviation of $5$. The Gaussian potential barrier as shown in (b) is centered at $ x_{b} = 80$ (red solid line) with width $ W = 1 $ (red dashed line represent $x_b\pm W$). Other parameter values are fixed to $H=1.5$, $R=1$, $X(0)=\sqrt{R-{1}/{\tau^2}}$, $Y(0)=0$, $Z(0)=0$.}
    \label{Fig:Prob evolution}
\end{figure}

\subsection{Evolution of probability density profiles}\label{Sec: Probab profiles}

For a quantum particle interacting with a finite potential barrier, one usually plots the evolution of the probability density profile as a function of time which clearly shows the transmitted and reflected part of the probability density after the initial distribution interacts with the barrier~\citep{griffiths_introduction_2018}. We create similar space-time plots showing the evolution of a distribution of initial conditions. The initial velocities of all WPEs is set to the steady-state positive velocity i.e. $X(0)=\sqrt{R-1/\tau^2}$ while $Y(0)=Z(0)=0$. The initial positions are sampled from a normal distribution with a mean of $60$ and a standard deviation of $5$. This results in an initial normal distribution of initial conditions in position and we plot its evolution in Fig.~\ref{Fig:Prob evolution}. In Fig.~\ref{Fig:Prob evolution}(a), for $\tau=1.5$, the probability density profile gets reflected from the potential barrier because the steady-state velocity is insufficient to cross the barrier. In Fig.~\ref{Fig:Prob evolution}(b), for 
$\tau=1.81$, even though the initial velocity is set at the steady state velocity, since $Y(0)$ and $Z(0)$ are not at its equilibrium values, we get velocity oscillations. Hence, when the probability density profile interacts with the barrier, part of it is transmitted and part of it is reflected due to the dynamical mechanism noted for Fig.~\ref{Fig:Steady Walking State2}. Further, the reflected and transmitted probability density profiles develop multiple peaks. We note that qualitative similar features of multiple peaks in the probability density have been observed in numerically simulated Bose-Einstein condensate (BEC) tunneling through a Gaussian potential barrier where both classical and quantum effects are present~\citep{BECtunneling}. Further increasing memory to $\tau=2$, the WPE velocity is sufficient for crossing the barrier (see Fig.~\ref{Fig:Steady Walking State3}(a)) and hence we see most of the probability density profile transmitting through the barrier. However, due to the presence of velocity oscillations, a qualitative change with multiple peaks is observed in the probability distribution after the transmission. Lastly, for $\tau=3.6$ in Fig.~\ref{Fig:Prob evolution}(d),  velocity oscillations are evident even before encountering the potential barrier, and after interaction with the barrier, transient chaos can occur as noted in Fig.~\ref{Fig:Steady Walking State3}(c)-(e). A part of the transmitted and reflected probability density profile corresponds to trajectories that experience little or no transient chaos, however, multiple peaks arise due to velocity oscillations. For the initial conditions undergoing transient chaos, they spend a long time in the vicinity of the barrier and then eventually get transmitted or reflected resulting in smaller probability density peaks transmitting or reflecting at later times. We note that in a recent work, \citet{Rahman2024} showed transmission and reflection of a probability density profile interacting with a rectangular potential barrier using their discrete-time model of walking droplets.

\section{Conclusions}\label{Sec: conclusion}

We have explored a dynamical analog of tunneling in a Lorenz-like model of an active WPE incident on a Gaussian potential barrier. We showed that at high memory, when the two symmetric fixed points of the Lorenz system become stable spirals, the WPE develops velocity oscillations and this results in alternating stripes in the basins of transmitted and reflected states in the initial condition space. This gives rise to sensitivity in barrier crossing based on initial conditions. Further, at even higher memories when transient chaos is present, pockets develop in the initial-condition space where the basin of the transmitted and reflected states are intermingled; this develops further sensitivity and unpredictability in barrier crossing based on initial conditions.

These dynamical features resulted in smooth variations of transmission probability as a function of system parameters. At low-memory just above the instability of stationary state, we found a discontinuous jump between no transmission and complete transmission as a function of dimensionless memory/height. Such a discontinuous jump in transmission probability would be expected for a classical particle crossing a potential barrier. At high memory, due to velocity fluctuations and transient chaos, this discontinuous jump became smooth with the transmission probability distribution taking intermediate values between zero and one. Such smooth variations in transmission probability arise for a quantum particle tunneling through a potential barrier. At very high memory, when the WPE motion became chaotic, further fluctuations develop in the transmission probability due to mixing of basins between transmitted and reflected states in the initial condition space. 

The evolution of a normally distributed position density profile constructed from ensemble of initial conditions showed transmitted and reflected probability profiles after interacting with the potential barrier. Further, velocity oscillations and transient chaos, induced wave-like features in the transmitted and reflected probability density profiles after interacting with the barrier. 

In the context of the experimental system of walking droplets, we note that our idealized model is not suitable for making quantitative predictions since it does not incorporate features of variations in vertical bouncing dynamics with memory~\citep{couchman_turton_bush_2019}, interactions with submerged barriers and wave modulation due to varying topography~\citep{Faria2017}. Moreover, the potential barrier in our model directly affects the droplet without affecting the droplet-generated waves. Conversely, in experiments of a walking droplet tunneling by \citet{Eddi2009} and \citet{tunneling2020}, the potential barrier is modeled as a submerged barrier whose effect on the droplet is mediated by the underlying waves. Hence, a more closer experimental setup to the model setup presented here would be to have a potential barrier in the form of an external magnetic field along with ferrofluid injected in the droplet, similar to the setup of \citet{Perrard2014a} created for generating a harmonic potential for the droplet. This method will ensure that the potential barrier is felt directly by the droplet without affecting the droplet-generated waves. Nevertheless, the key non-trivial qualitative features observed in this work, such as the stripes and intermingled patterns in the initial condition space and the enhanced sensitivity to barrier crossing, emerge from unsteady motion associated with velocity oscillations. Since the dynamical regime of a free walker undergoing velocity oscillations has been demonstrated experimentally~\citep{Bacot2019}, it would be interesting to revisit the experiments of a walking droplet tunneling by \citet{Eddi2009} and \citet{tunneling2020} in the high memory regime, or perform experiments with a magnetic field potential barrier in the high memory regime, when the free walker is undergoing velocity fluctuations before encountering the barrier. As observed in our simple model, this might result in further sensitivity to barrier crossing which would now be rooted in sensitivity to initial conditions due to unsteady dynamical features. Lastly, in the context of generalized pilot-wave framework~\citep{Bush2015}, our work demonstrates that nonlinear and chaotic features of a simple classical pilot-wave system provide a dynamical mechanism for the origin of sensitivity and unpredictability in barrier crossing.


\begin{acknowledgments}
R.X. was partly supported by Adelaide Summer Research Scholarship (ASRS) awarded by the University of Adelaide. R.V. was supported by Australian Research Council (ARC) Discovery Project DP200100834 and by the Leverhulme Trust [Grant No. LIP-2020-014] during the course of the work. Some of the numerical results were computed using supercomputing resources provided by the Phoenix HPC service at the University of Adelaide and the Hydra cluster in the Department of Physics at the University of Oxford.
\end{acknowledgments}



\bibliography{TunnelingWPE}

\end{document}